\documentclass[prl,twocolumn,superscriptaddress,amsmath,amsthm,showpacs]{revtex4}
\usepackage{amssymb}
\usepackage{color}
\usepackage{graphicx}

\newcommand{\bra}[1]{\langle #1 |}
\newcommand{\ket}[1]{| #1 \rangle}
\newcommand{\braket}[2]{\langle #1 | #2 \rangle}
\newcommand{\ketbra}[2]{| #1 \rangle \langle #2 |}
\newcommand{\expect}[1]{\langle #1 \rangle}

\begin{document}
\title{Testing nonclassicality and non-Gaussianity in phase space}
\author{Jiyong Park$^\dag$}
\affiliation{Department of Physics, Texas A\&M University at Qatar, Education City, P.O.Box 23874, Doha, Qatar}
\author{Junhua Zhang$^\ddag$}
\affiliation{Center for Quantum Information, Institute for Interdisciplinary Information Sciences, Tsinghua University, Beijing 100084, People’s Republic of China}
\author{Jaehak Lee}
\affiliation{Department of Physics, Texas A\&M University at Qatar, Education City, P.O.Box 23874, Doha, Qatar}
\author{Se-Wan Ji}
\affiliation{Department of Physics, Texas A\&M University at Qatar, Education City, P.O.Box 23874, Doha, Qatar}
\author{Mark Um}
\affiliation{Center for Quantum Information, Institute for Interdisciplinary Information Sciences, Tsinghua University, Beijing 100084, People’s Republic of China}
\author{Dingshun Lv}
\affiliation{Center for Quantum Information, Institute for Interdisciplinary Information Sciences, Tsinghua University, Beijing 100084, People’s Republic of China}
\author{Kihwan Kim*}
\affiliation{Center for Quantum Information, Institute for Interdisciplinary Information Sciences, Tsinghua University, Beijing 100084, People’s Republic of China}
\author{Hyunchul Nha$^\P$}
\affiliation{Department of Physics, Texas A\&M University at Qatar, Education City, P.O.Box 23874, Doha, Qatar}
\affiliation{School of Computational Sciences, Korea Institute for Advanced Study, Seoul 130-722, Korea}
\date{\today}

\begin{abstract}
We theoretically propose and experimentally demonstrate a nonclassicality test of single-mode field in phase space, which has an analogy with the nonlocality test proposed by Banaszek and W{\'o}dkiewicz [Phys. Rev. Lett. {\bf 82}, 2009 (1999)]. Our approach to deriving the classical bound draws on the fact that the Wigner function of a coherent state is a product of two independent distributions as if the orthogonal quadratures (position and momentum) in phase space behave as local realistic variables. Our method detects every pure nonclassical Gaussian state, which can also be extended to mixed states. Furthermore, it sets a bound for all Gaussian states and their mixtures, thereby providing a criterion to detect a genuine quantum non-Gaussian state. Remarkably, our phase-space approach with invariance under Gaussian unitary operations leads to an optimized test for a given non-Gaussian state. We experimentally show how this enhanced method can manifest quantum non-Gaussianity of a state by simply choosing phase-space points appropriately, which is essentially equivalent to implementing a squeezing operation on a given state. 
\end{abstract}

\maketitle

{\it Introduction}---Nonclassicality of a quantum state is a topic of crucial importance that has attracted a lot of theoretical and experimental efforts for long. It provides not only a profound conceptual framework to distinguish quantum phenomena from classical ones, but also an essential practical basis for numerous applications, e.g. in quantum information processing. An important approach to studying quantum mechanics in comparison with classical mechanics is to adopt a phase-space description of a quantum state \cite{Wigner1932}. A wide variety of quantum systems of continuous variables (CVs) can be addressed in phase space, including quadrature amplitudes of light fields, collective spin states of atomic ensembles, and motional states of trapped ions, Bose-Einstein condensate, or mechanical oscillators, etc. \cite{Cerf}. Investigating quantum dynamics in phase space has yielded a great deal of intuition to quantum-to-classical transition \cite{Zurek2003}. It also offers a powerful tool to treat problems in quantum optics \cite{Barnett} and CV quantum informatics \cite{Braunstein2005}. 

A clear signature of nonclassicality is the negativity of phase-space distribution, which does not exist in classical probability distributions. However, its demonstration  typically requires a full reconstruction of Wigner function \cite{Lvovsky2009} and it is of fundamental and practical significance to have a simpler set of measurements manifesting nonclassicality \cite{Richter,Mari}, desirably even when the Wigner function is non-negative. 
For instance, every Gaussian state possesses a positive-definite Wigner function, which restricts a possible set of nonclassicality tests. To manifest the Bell nonlocality, e.g., by employing homodyne detections, a Gaussian state must be transformed to a non-Gaussian state having a non-positive Wigner function to rule out hidden-variable models \cite{Bell, Nha}. Banaszek and W{\'o}dkiewicz (BW) introduced a different seminal approach to manifesting CV nonlocality even with a positive-definite Wigner function \cite{Banaszek1999}. Their method looks into the particle nature of the field, i.e. the photon-number parity directly related to Wigner function, in contrast to the wave nature revealed by homodyne detections. The BW formalism has significantly contributed to a deeper understanding of nonclassical correlation, which has also been extended to multipartite systems, e.g. \cite{Kim, Adesso2014}. Despite its intellectual impact, the BW test was not yet experimentally achieved, furthermore, a similar phase-space formalism to demonstrate nonclassicality at a more elementary single-mode level was not developed so far.

Here we propose a nonclassicality test of single-mode fields in phase space. Our nonclassicality refers to those states that cannot be represented as a mixture of coherent states. We first note that the Wigner function of a coherent state is a product of two Gaussian distributions, leading to a notion of classicality that two orthogonal quadratures in phase space behave as local realistic variables. This makes a nonclassicality test, in analogy with the BW formalism, taking four points in phase space. Our test can detect every pure Gaussian nonclassical state. Furthermore, it sets a bound for all Gaussian states and their mixtures, thus offering a test of quantum non-Gaussianity---a stronger form of nonclassicality. Numerous applications require non-Gaussian resources as essential, e.g. universal CV quantum computation \cite{Lloyd}, entanglement distillation \cite{Eisert}, quantum error correction \cite{Niset}, and CV nonlocality test \cite{Nha}, etc.. It is thus of crucial importance to distinguish a genuinely quantum non-Gaussian state from a mixture of only Gaussian states \cite{Filip2014,negativity}. Note that the latter can also be a non-Gaussian state when a finite sum of Gaussian states is taken, giving a nonzero value of some known non-Gaussianity measures \cite{Genoni2007}. However, it cannot be regarded as a genuine non-Gaussian resource. Remarkably, we present a method to optimize the non-Gaussianity test using an invariance property under Gaussian unitary operations. It enables us to detect a broad set of non-Gaussian states, which is experimentally demonstrated to illustrate the power of our approach.

	\begin{figure}[!]
		\includegraphics[scale=0.32]{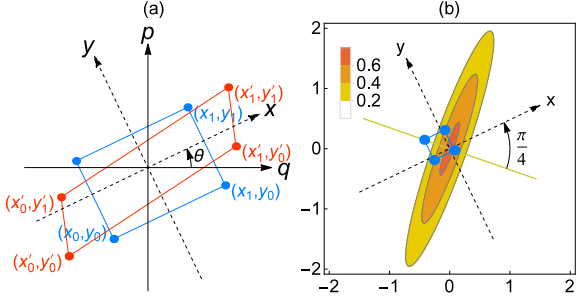}
		\caption{(a) Four phase-space points chosen at the vertices of rectangle for a nonclassicality test (blue solid) and of parallelogram for a non-Gaussianity test (red solid) (b) An optimal choice of rectangle for testing a squeezed state occurs at $\theta -\phi= \frac{\pi}{4}$  ($\phi$: squeezing axis) .}
		\label{fig:R}
	\end{figure}

{\it Nonclassicality}---The Wigner function $W_{\rho} ( q, p )$ of a state $\rho$ is directly related to the number parity of its displaced state $\tilde\rho_\alpha\equiv\hat{D} (\alpha )\rho\hat{D}^{\dag} (\alpha )$, where $\hat{D} ( \alpha ) = e^{\alpha \hat{a}^{\dag} - \alpha^{*} \hat{a}}$ displaces the given state by amplitude $\alpha\equiv q+ip$ \cite{Barnett}. That is, 
\begin{eqnarray}
\frac{\pi}{2} W_{\rho} ( q, p ) = \mathrm{tr} [ (-1)^{\hat{a}^{\dag} \hat{a}}\tilde\rho_\alpha],
\end{eqnarray} thus taking a value in [-1,1]. Notably, a coherent state $\ket{\alpha}$ has a symmetric Wigner function that is a product of two Gaussian distributions for any pair of orthogonal quadratures,
$	\frac{\pi}{2}W_{\ketbra{\alpha}{\alpha}} ( x, y ) = e^{- 2 ( x - \alpha_{x} )^{2}} e^{- 2 ( y - \alpha_{y} )^{2}},$
where $( x, y )^{T} = \mathcal{R} ( \theta ) ( q, p )^{T}$ is the coordinate system  rotated by an angle $\theta$ from $(q,p)$ in phase space (Fig.~1 (a)), with $\alpha_{x} = \mathrm{Re} [ \alpha e^{-i \theta} ]$ and $\alpha_{y} = \mathrm{Im} [ \alpha e^{-i \theta} ]$. 
The Wigner function of a coherent state can thus be regarded as a product of two independent random variables, $\frac{\pi}{2} W_{\rho} ( q, p )=ab$ ($0\le a,b\le1$). 

Let us define \begin{equation} \label{eq:nonclassicality}
		\mathcal{J} [ \rho ] \equiv \frac{\pi}{2} \sum_{j = 0}^{1} \sum_{k = 0}^{1} (-1)^{jk} W_{\rho} ( x_{j}, y_{k} ),
	\end{equation}
where the four points $( x_{j}, y_{k} )$ form a rectangle oriented at angle $\theta$ in phase space (Fig.~\ref{fig:R} (a)). We then have $\mathcal{J} [ \rho ]=a_0b_0+a_1b_0+a_0b_1-a_1b_1$ for a coherent state, which is the very structure of the celebrated CHSH inequality \cite{CHSH}. Due to the condition $0\le a,b\le1$, we obtain $-1 \leq \mathcal{J} \leq 2$ \cite{note}. Extending to a mixture of coherent states, $\rho=\int p ( \lambda ) \ketbra{\lambda}{\lambda}$, we derive a classicality condition as 
\begin{eqnarray}
-1 \leq \mathcal{J} [ \rho_{\mathrm{cl}} ] \leq 2.
\end{eqnarray}
 We note that $	\frac{\pi}{2}W_{\int p ( \lambda ) \ketbra{\lambda}{\lambda}} ( x, y ) = \int p ( \lambda ) a ( x|\lambda ) b ( y|\lambda )$, where the coherent amplitude $\lambda$ plays a role as a hidden variable. Violation of inequality (3) indicates a nonclassicality of single-mode fields.
  
Our nonclassicality test bears a close connection to the BW formalism---a CHSH-type nonlocality test in phase space directly addressing two systems \cite{Banaszek1999}---particularly in obtaining the classical upper bound 2. On the other hand, our test has a different lower bound -1, of which violation also indicates nonclassicality. This necessarily requires the negativity of the Wigner function and we focus on the violation of the upper bound in this Letter \cite{Supple}. Furthermore, unlike the CHSH inequality allowing a maximum $2\sqrt{2}$ (Cirelson's bound) \cite{Cirelson1980}, the maximum value of $\mathcal{J}$ in our test approaches 4, which can be attributed to the noncommutativity of the local operators $\hat{q},\hat{p}$ and $\hat{n}$ \cite{Supple}. 


	\begin{figure}[!]
		\includegraphics[scale=0.4]{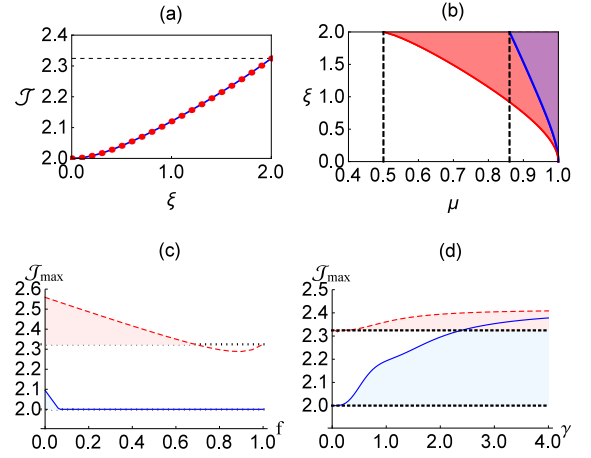}
		\caption{(a) Optimal $\mathcal{J}$ for a pure Gaussian state $\sigma_{\rm G}$ against squeezing $\xi = 2 \tanh 2r$ up to $\mathcal{B}_{G} = \frac{8}{3^{9/8}} \approx 2.32$ (gray dashed line). (b) Detection of nonclassicality for Gaussian states against purity $\mu$ and squeezing parameter $\xi$. Our test adopting four (three) phase-space points detects nonclassicality with purity $\mu>\frac{3^{9/8}}{4}\approx0.86$ ($\mu>0.5$) represented by a purple (red) shaded region. 
		(c) Maximal $\mathcal{J}$ for $f \ketbra{0}{0} + (1-f) \ketbra{2}{2}$ and (d) $e^{- \gamma^{2}} ( \sinh \gamma^{2} \ketbra{0}{0} + \cosh \gamma^{2} \ketbra{\psi_{\gamma}}{\psi_{\gamma}} )$, using a rectangle (blue solid line) and  a parallelogram (red dashed line) test by optimizing phase-space points for each state. $\mathcal{J}>8/3^{9/8}\approx2.32$ and $\mathcal{J}>2$ represent quantum non-Gaussianity and nonclassicality, respectively. The Wigner function becomes positive for $\frac{1}{2} < f \leq 1$ and $0 < \gamma$ in (c) and (d), respectively.}
		\label{fig:OP}
	\end{figure}

{\it Gaussian states}---We first show how our test detects a Gaussian nonclassical state. A single-mode Gaussian state can generally be represented as a displaced squeezed thermal state,
	\begin{equation}
		\sigma_{\rm G} = \hat{D} ( \alpha ) \hat{S} ( r, \phi ) \sigma_{\rm th} ( \bar{n} ) \hat{S}^{\dag} ( r, \phi ) \hat{D}^{\dag} ( \alpha),
	\end{equation}
where $\hat{S} ( r, \phi ) = \exp [ - \frac{r}{2} ( e^{2 i \phi} \hat{a}^{\dag 2} - e^{- 2 i \phi} \hat{a}^{2} ) ]$ is the squeezing operator ($r$: squeezing strength, $\phi$: squeezing axis), and $\sigma_{\rm th} ( \bar{n} )= \sum_{n = 0}^{\infty} \frac{\bar{n}^{n}}{( \bar{n} + 1 )^{n + 1}} \ketbra{n}{n}$ is a thermal state ($\bar{n}$: mean photon number). 
For a given Gaussian state with $\{ \alpha, r, \phi, \bar{n} \}$, an optimal violation occurs at four vertices of a rectangle oriented $45^{\rm o}$ from the squeezed axis, i.e. $\theta -\phi= \frac{\pi}{4}$ [Fig. 1 (b)] \cite{Supple}. Intuitively, this choice of rectangle makes a sense because our test of nonclassicality relies on the non-factorizability of Wigner function (See the paragraphs below Eq. (1)). The optimal points are 
	\begin{align} \label{eq:parameters}
		x_{j} -\alpha_{x} = (-1)^{j} z_{j} \kappa, \hspace{0.2in}
		y_{j} -\alpha_{y}= (-1)^{j+1} z_{j} \kappa,  
	\end{align}
$(j=0,1)$, with $\kappa = \sqrt{\frac{2 \bar{n} + 1}{2 \cosh 2r}}$ and $z_{j}$ given in \cite{Supple}.

In Fig.~\ref{fig:OP} (a), we plot an optimal $\mathcal{J} [ \sigma_{\rm G}]$ for a pure Gaussian state against the squeezing strength $\xi = 2 \tanh 2r$, whose values are all above 2 demonstrating a successful detection of nonclssicality. To fully identify the detectable regime including mixed Gaussian states, we draw a contour plot of $\mathcal{J} [ \sigma_{\rm G}]$ against purity $\mu = \frac{1}{2 \bar{n} + 1}$ and squeezing $\xi$ in Fig.~\ref{fig:OP} (b). 
Our test detects a mixed Gaussian state above the purity $\mu>\frac{3^{9/8}}{4}\approx0.86$ (purple shaded region). Remarkably, the detection can be further enhanced to $\mu>0.5$ (red shaded region) by a simpler test of adopting three phase-space points that has no analogue in BW inequality \cite{Supple}.

Among all Gaussian states, we analytically obtain the maximal value $\mathcal{B}_{G} = \frac{8}{3^{9/8}}\approx 2.32$ in the limit $\xi \rightarrow 2$ (infinite squeezing), at points $( x_{0}, y_{0}, x_{1}, y_{1} ) = ( \kappa t, -\kappa t, -3\kappa t, 3\kappa t )$ with $t = \frac{1}{4} \sqrt{\frac{\log 3}{2}}$ \cite{Supple}. This value coincides with the maximal Gaussian bound for two-mode \cite{Jeong2003} and three-mode \cite{Adesso2014} nonlocality tests in phase space.

{\it Genuine quantum non-Gaussianity}---A non-Gaussian state can take a larger value than the Gaussian bound, $\mathcal{J}>\mathcal{B}_{\mathrm{G}}$, which becomes a sufficient condition to detect quantum non-Gaussianity. As the maximum achieved by a Gaussian state $\sigma_G$ is $\mathcal{B}_{\mathrm{G}}$, the same bound also applies to a mixture of Gaussian states, $\rho_{\mathrm{MG}}=\sum_ip_i\sigma_G^i$. Importantly, under an arbitrary Gaussian unitary operation $\hat{U}_{G}$, the Gaussian mixture $\rho_{\mathrm{MG}}$ remains as a Gaussian mixture, thus $\mathcal{J} [ \hat{U}_{G} \rho_{\mathrm{MG}} \hat{U}_{G}^\dag ] \leq \mathcal{B}_{\mathrm{G}}$, which enables us to construct an optimized test of non-Gaussianity. 
\\{\bf Criterion}: {\it Given a state $\rho$, if there exists a $\hat{U}_{G}$ making
	\begin{equation}
		\mathcal{J} [ \hat{U_{G}} \rho \hat{U}^{\dag}_{G} ]> \mathcal{B}_{\mathrm{G}}=8/3^{9/8},
		\label{eqn:criterion}
	\end{equation}
the state $\rho$ is genuinely quantum non-Gaussian.}

	
$\hat{U}_{G}$ refers to displacement, phase-shift, and squeezing operations, however, our test does not require the implementation of those operations. 
First, displacement and phase-shift do not affect an optimal $\mathcal{J} [\rho]$, thus excluded in our consideration. They only shift or rotate the profile of Wigner function while keeping its entire shape, thus the optimal points are simply moved along the distribution but yielding the same maximum \cite{Supple}. On the other hand,  
under a squeezing operation, 
 using  Eq.~(1) and the identity ${\hat S}^\dag(r,\phi){\hat D}(\alpha){\hat S}(r,\phi)={\hat D}({\tilde \alpha})$, where ${\tilde \alpha}=\alpha\cosh r+\alpha^*e^{2i\phi}\sinh r$, we obtain 
$W_{\hat{S} \rho \hat{S}^{\dag}}( x, y )=W_\rho( \tilde{x}, \tilde{y})$, with
$( \widetilde{x}, \widetilde{y})^{T} = \mathcal{S} ( r,\phi  ) ( x, y )^{T}$. The linear transformation $\mathcal{S} ( r,\phi )=\cosh r I+\sinh r\cos 2\phi\sigma_z+\sinh r\sin 2\phi\sigma_x$ ($\sigma_i$: Pauli matrix) generally shifts four points at the vertices of rectangle to those of parallelogram (Fig.~1 (a)). Therefore, we can essentially implement the ``rectangle" test for the squeezed state $\hat{S} \rho \hat{S}^{\dag}$ by simply considering four vertices of the corresponding parallelogram for the given state $\rho$, i.e. $
		\mathcal{J} [\hat{S} \rho \hat{S}^{\dag}] = \frac{\pi}{2} \sum_{j,k = 0}^{1} ( - 1 )^{jk} W_{\rho} ( \tilde{x}_j, \tilde{y}_k )$.

We illustrate the power of our approach by considering some examples, compared with other known criteria based on the Wigner- and the Q-function under energy constraint in \cite{Genoni2013, Hughes2014}, respectively. For a mixture of vacuum and two photon states $f \ketbra{0}{0} + (1-f) \ketbra{2}{2}$, our criterion detects quantum non-Gaussianity for $0 \leq f \lesssim 0.69$ as shown in Fig. 2 (c), whereas the methods of \cite{Genoni2013} and \cite{Hughes2014} detect none and for $0 \leq f \lesssim 0.093$, respectively.
The Wigner function of a Fock state $\ket{n}$ is given by
$W_{\ketbra{n}{n}} ( q, p ) = \frac{2}{\pi} (-1)^{n} e^{- 2 q^{2} - 2 p^{2}} L_{n} ( 4 q^{2} + 4 p^{2} ), $
where $L_{n}$ is the Laguerre polynomial of order $n$. 
For the mixture $f \ketbra{0}{0} + (1-f) \ketbra{2}{2}$, the Wigner function becomes positive definite for $\frac{1}{2} \leq f \leq 1$. Importantly, Fig. 2 (c) shows how our detection can be significantly enhanced by implementing the parallelogram test (red dashed line) in contrast to the rectangle test (blue solid line). Similarly, our test detects a mixture of even cat state $\ket{\psi_{\gamma}} = \mathcal{N} ( \ket{\gamma} + \ket{- \gamma} )$ with coherent amplitude $\gamma$ and vacuum, i.e. $e^{- \gamma^{2}} ( \sinh \gamma^{2} \ketbra{0}{0} + \cosh \gamma^{2} \ketbra{\psi_{\gamma}}{\psi_{\gamma}} )$ for all non-zero $\gamma$ (Fig.~2 (d)), whereas the witnesses in \cite{Genoni2013, Hughes2014} detect none of these states. 

	
	\begin{figure}
		\includegraphics[scale=0.4]{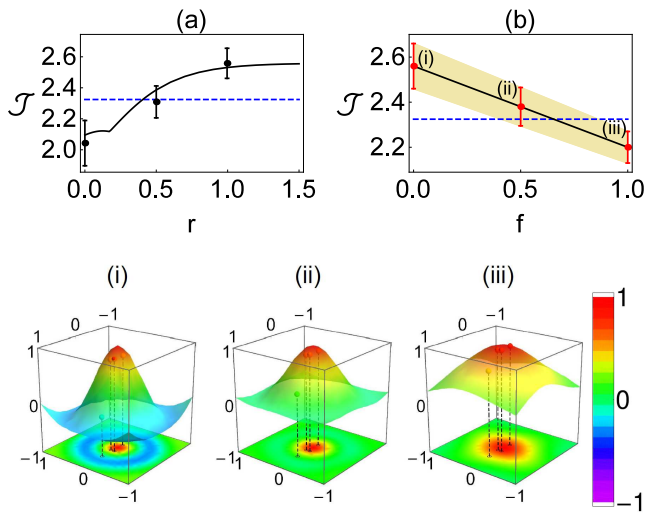}
		\caption{(Color online)  (a) $\mathcal{J}$ for a Fock state $\ket{2}$ by using a parallelogram test [Fig. 1 (a)] that essentially implements squeezing ($r$: strength) on a given state. Theoretical predictions (black solid line) are in good agreement with the experimental data (filled circles). Genuine non-Gaussianity is manifested above $\mathcal{B}_{\mathrm{G}}=8/3^{9/8}$ (blue dashed line). (b) $\mathcal{J}$ for $f \ketbra{0}{0} + (1-f) \ketbra{2}{2}$ measured in experiment using the parallelogram test with the fixed set of four phase-space points, ( -0.110, -0.110 ), ( 0.121, 0.100 ), ( 0.100, 0.121 ), and ( 0.331, 0.331 ). 
		Red circles (i), (ii), and (iii) are results for states $\ketbra{2}{2}$, $(\ketbra{0}{0}+\ketbra{2}{2})/2$, and $\ketbra{0}{0}$, respectively, with their experimentally constructed Wigner functions shown below. } 
		\label{fig:VF2}
	\end{figure}

{\it Experiment}---
We have performed an experiment for the state $f \ketbra{0}{0} + (1-f) \ketbra{2}{2}$ using a single motional mode ($\rm X$-direction) of a trapped $^{171}$Yb$^{+}$ ion in 3-dimensional harmonic potential with trap frequencies $\left(\omega_{\rm X},\omega_{\rm Y}, \omega_{\rm Z}\right) = 2 \pi ~ (2.8,3.2,0.6)$ MHz. We employ the coupling of phonon mode to two electric levels of $^{171}$Yb$^{+}$ ion in the $S_{1/2}$ ground state manifold, $\ket{F=1,m_{F}=0}\equiv\ket{\uparrow}$ and $\ket{F=0,m_{F}=0}\equiv\ket{\downarrow}$ with separation $\omega_{\rm HF} = (2 \pi) ~ 12.642821$ GHz. The control and manipulation of phonon mode $\hat a$ is made by the anti-Jaynes-Cumming interaction $H_{\rm aJC} = \frac{\eta \Omega}{2} \hat{a}^{\dagger} \hat{\sigma}_{+} + {\rm h.c.}$ with $\sigma_{+}=\ket{\uparrow}\bra{\downarrow}$, realized by two counter-propagating laser beams with beat frequency near $\omega_{\rm HF} + \omega_{\rm X}$ \cite{Supple}.  Here $\eta = \Delta k \sqrt{\hbar/2 M \omega_{\rm X} }$ is the Lamb-Dicke parameter, $\Omega$ the Rabi frequency of qubit transition, $\Delta k$ the net wave-vector of the Raman laser beams and $M$ the ion mass. 

To test our criterion, we need to (i) prepare an initial state, (ii) displace it, and (iii) measure the parity, sequentially. 
(i) We prepare the motional ground state $\ket{n=0}$ by Doppler cooling and the standard Raman sideband cooling. For the preparation of $\ket{n=2}$ state, we successively apply a $\pi$-pulse of $H_{\rm aJC}$ transferring the state $\ket{\downarrow, n}$ to $\ket{\uparrow, n+1}$, and a $\pi$-pulse for internal state transition $\ket{\uparrow, n+1}$ to $\ket{\downarrow, n+1}$.
(ii) We then displace the initial state $\ket{n=0}$ or $\ket{n=2}$.
We perform a displacement operation $\hat{D}(\alpha)$ using the Hamiltonian $H_{D} = \frac{\eta\Omega_{D}e^{i \phi}}{2} \hat{a}^{\dagger} + {\rm h.c.}$, realized by the same Raman laser beams with the beat frequency near $\omega_{\rm X}$ and phase $\phi$, where the displacement is given by $\alpha = -i\frac{\eta\Omega_{D} e^{i \phi}}{2}t$. (See \cite{Supple} and \cite{KKim2014} for further description of experimental setting.)  
(iii) After each displacement $\hat{D}(\alpha)$, we obtain the population $Q_{n}(\alpha)= \bra{n} \hat{D}(\alpha) \rho  \hat{D}^{\dagger}(\alpha) \ket{n}$ of Fock state $\ket{n}$ by observing  $\ket{\uparrow}$-state probability under the evolution with $H_{\rm aJC}$ and fitting the signal to \cite{WWineland96}
\begin{eqnarray}
P_{\uparrow}\left(t\right)=\frac{1}{2}\sum_{n} Q_{n}(\alpha)\left(1-A_{b}e^{-\lambda_{b}t}\cos\left(\Omega_{n,n+1}t\right)\right).
\label{eq:fit}   
\end{eqnarray}
$\Omega_{n,n+1}\approx \sqrt{n+1} \eta \Omega $ is the Rabi oscillation rate between $\ket{\downarrow,n}$ and $\ket{\uparrow,n+1}$ while $A_{b}$ and $\lambda_{b}$ are parameters incorporating experimental imperfections. We obtain the value of Wigner function $\frac{\pi}{2} W(x,y)$ by $\frac{\pi}{2} \sum_{n} (-1)^{n} Q_{n}(x,y)$. We observed the evolution with $H_{\rm aJC}$ in steps of 1 $\mu$s for a total duration of $150 ~\mu$s, about 20 times the $\pi$-pulse duration of $\Omega_{0,1}$, with 100 measurements at each step \cite{Supple}. We repeated experiments 10 times to obtain the parity under each displacement. 

We first perform an experiment on the state $\ket{n=2}$ by changing the shape of parallelogram in phase space (Fig. 1 (a)), essentially the same as applying squeezing operation in Eq. (\ref{eqn:criterion}). The experimental data in Fig. 3 (a) confirms the increasing capability of our test with squeezing to manifest genuine non-Gaussianity. 

In a second experiment, we obtained $\mathcal{J}=2.20 \pm 0.07$ for the ground state $\ket{n=0}$ and $\mathcal{J}=2.56 \pm 0.1$ for the state $\ket{n=2}$, with the same set of four phase-space points [Fig. 3 (b)] \cite{nha1}. A mixture of these states $f \ketbra{0}{0} + (1-f) \ketbra{2}{2}$ can be addressed by including all data together for both states with an appropriate weighting \cite{nha2}. We observed $\mathcal{J}>\mathcal{B}_{\mathrm{G}}=8/3^{9/8}$ even for $0.5\le f\le0.65$, where no negativity appears in the Wigner function (See \cite{Supple} for the experimental construction of the Wigner function.) Due to the fluctuations of parameters in experimental system, we have standard deviation of parity about 5 $\%$ of the mean values, which could be further reduced by repeating experiments more.  

In summary, we developed a phase-space approach to manifest nonclassicality of a single-mode quantum state. Our approach has a close connection to the nonlocality test ruling out local hidden-variable models, particularly the BW-CHSH inequality, using that two quadratures of a coherent state behave like local variables in phase space. Our test was further extended to detection of genuine quantum non-Gaussianity, which is known to be an essential resource for many applications. Our methods successfully detect nonclassicality and non-Gaussinity for a broad class of Gaussian and non-Gaussian states even when the Wigner functions are positive definite, which was further confirmed by experimental data. Our phase space approach for single-mode fields and the BW approach for multi-mode fields can therefore be crucial tools together to shed light on nonclassicality of CV systems.

While we demonstrated our criterion using a trapped ion system, we anticipate that other CV systems can also be experimentally addressed, e.g. quantum optical systems \cite{Laiho2010,Silberhorn}. For example, Harder {\it et al.} recently showed an experimental method of obtaining photon-number distribution robust against detection inefficiency by employing only a thermal state as a probe and an on-off detector \cite{Silberhorn}. Combined with several displacement operations, it can be directly used to test our proposed criteria. 

{\it Acknowledgements}---This work was supported by an NPRP grant 4-346-1-061 from Qatar National Research Fund and by the National Basic Research Program of China under Grants No. 2011CBA00300 (No. 2011CBA00301), the National Natural Science Foundation of China 61073174, 61033001, 61061130540, and 11374178. K.K. acknowledges the first recruitment program of global youth experts of China.

\hspace{-0.35cm}
$\dag$ jiyong.park@qatar.tamu.edu, $\ddag$ zjh26108@gmail.com, 
* kimkihwan@gmail.com\\
$\P$ corresponding author: hyunchul.nha@qatar.tamu.edu

\newpage

\hspace{1cm}{\bf SUPPLEMENTAL MATERIAL}

\section{A. Testing Gaussian states}
{\bf (i) Representing an arbitrary Gaussian state}\\
A Gaussian state $\sigma$ is fully characterized by its first-order moments (averages) $\langle\hat{q}\rangle$ and $\langle\hat{p}\rangle$ and second-order moments (covariance matrix elements) 
\begin{equation}
		\Gamma_{jk} = \frac{1}{2} \expect{\hat{R}_{j} \hat{R}_{k} + \hat{R}_{j} \hat{R}_{k}} - \expect{\hat{R}_{j}} \expect{\hat{R}_{k}},\hspace{0.3cm} j,k=1,2
	\end{equation}
where $\hat{R} \equiv ( \hat{q}, \hat{p} )$ with $\hat{q} = \frac{1}{2} ( \hat{a}^{\dag} + \hat{a} )$ and $\hat{p} = \frac{1}{2i} ( \hat{a}-\hat{a}^{\dag} )$. 
An arbitrary single-mode Gaussian state $\sigma = \hat{D} ( \alpha ) \hat{S} ( r, \phi ) \sigma_{th} ( \bar{n} ) \hat{S}^{\dag} ( r, \phi ) \hat{D}^{\dag} ( \alpha)$ introduced in the main Letter has the averages $\langle\hat{q}\rangle=\mathrm{Re} [ \alpha ]$, $\langle\hat{p}\rangle=\mathrm{Im} [ \alpha ]$, and
the covariance matrix elements 
	\begin{align} \label{eq:displaced-squeezed-thermal-state}
		\Gamma_{11} & = \frac{1}{2}( \bar{n} + {\textstyle \frac{1}{2}} ) ( \cosh 2r - \sinh 2r \cos 2 \phi ), \nonumber \\
		\Gamma_{22} & = \frac{1}{2}( \bar{n} + {\textstyle \frac{1}{2}} ) ( \cosh 2r + \sinh 2r \cos 2 \phi ), \nonumber \\
		\Gamma_{12} & =\Gamma_{21}= - \frac{1}{2}( \bar{n} + {\textstyle \frac{1}{2}} ) \sinh 2r \sin 2 \phi.
	\end{align}
Its Wigner function takes a Gaussian form as
	\begin{equation} \label{eq:smWigner1}
		W_{\sigma} ( q, p ) = \frac{2 \mu}{\pi} \exp [ - \frac{1}{2}( Q - \expect{Q} )^{T} \Gamma^{-1} ( Q - \expect{Q} ) ],
	\end{equation}
where $Q = ( q, p )^{T}$, $\expect{Q} = ( \mathrm{Re} [ \alpha ], \mathrm{Im} [ \alpha ] )^{T}$,  and $\mu = \frac{1}{4 \sqrt{\det \Gamma}}=\frac{1}{1+2\bar{n}}$ is the purity of the state. 

If we rewrite Eq.~\eqref{eq:smWigner1} using the rotated quadratures $\widetilde{Q} = (x,y)^{T} = \mathcal{R} ( \theta ) (q,p)^{T}$ (Fig. 1 (a) in the main Letter), we have
	\begin{equation} \label{eq:smWigner2}
		W_{\sigma} ( x, y ) = \frac{2 \mu}{\pi} \exp [ -\frac{1}{2}( \widetilde{Q} - \expect{\widetilde{Q}} )^{T} \widetilde{\Gamma}^{-1} ( \widetilde{Q} - \expect{\widetilde{Q}} ) ],
	\end{equation}
where $\widetilde{\Gamma} = \mathcal{R} ( \theta ) \Gamma \mathcal{R} ( - \theta )$ represents the covariance matrix in a rotated frame, with
	\begin{equation}
		\mathcal{R} ( \theta ) =
		\begin{pmatrix}
			\cos \theta & \sin \theta \\
			- \sin \theta & \cos \theta
		\end{pmatrix}, 
	\end{equation}
and $\expect{\widetilde{Q}} = ( \mathrm{Re} [ \alpha e^{-i \theta} ], \mathrm{Im} [ \alpha e^{-i \theta} ] )$. Our task is then to optimize $\mathcal{J}$ in Eq. (2) of the main Letter by varying the angle $\theta$ in choosing a rotated rectangle and the coordinates $\{x_0,y_0,x_1,y_1\}$ of its four vertices.

{\bf (ii) Invariance of maximal $\mathcal{J}$ under displacement and phase-shift operations}\\
We first note that the maximal value $\mathcal{J}$ for a given state $\rho$ (both Gaussian and non-Gaussian) is the same as that of its displaced state $\hat{D} ( \alpha ) \rho \hat{D}^{\dag} ( \alpha )$: If the maximum for $\rho$ occurs at points $\{x_0,y_0,x_1,y_1\}$, the same maximum then occurs for the displaced state at shifted points $( x', y' ) =( x + \mathrm{Re} [ \alpha e^{-i \theta} ], y + \mathrm{Im} [ \alpha e^{-i \theta} ] )$, because the displacement only shifts the center of the Wigner distribution, while keeping its entire profile. We can thus ignore the first-order moments of a given state to find its maximum $\mathcal{J}$. Secondly, a phase-shift operation does not change the maximum $\mathcal{J}$, either. If the maximum for $\rho$ occurs at $\{x_0,y_0,x_1,y_1\}$ of four vertices of rectangle oriented at angle $\theta$, the same maximum for the rotated state $e^{i\varphi a^\dag a}\rho e^{-i\varphi a^\dag a}$ occurs at the vertices of the identical rectangle oriented now at angle $\theta+\varphi$, because the phase-rotation also keeps the profile of Wigner function.

{\bf (iii) Maximum $\mathcal{J}$ for a Gaussian state}\\
Therefore, we set $\alpha = 0$ to find a maximum $\mathcal{J}$ without loss of generality. Introducing rescaled coordinates 
\begin{eqnarray}
( x^{\prime}, y^{\prime} ) = 2 \mu ( x \sqrt{2\widetilde{\Gamma}_{22}}, y \sqrt{2\widetilde{\Gamma}_{11}} ),
\end{eqnarray} 
we recast Eq.~\eqref{eq:smWigner2} to
	\begin{equation}
		W_{\sigma} ( x^{\prime}, y^{\prime} ) = \frac{2 \mu}{\pi} \exp [ - x^{\prime 2} - y^{\prime 2} + k x^{\prime} y^{\prime} ],
	\end{equation}
with
	\begin{equation}
		k = \frac{2 \widetilde{\Gamma}_{12}}{\sqrt{\widetilde{\Gamma}_{11} \widetilde{\Gamma}_{22}}} = \frac{2 \tanh 2r \sin 2 ( \theta-\phi )}{\sqrt{1 - \tanh^{2} 2r \cos^{2} 2 ( \theta-\phi )}}.
	\end{equation}
The purity $\mu=\frac{1}{1+2\bar{n}}$ only affects the overall factor, with $k$ independent of it. We thus set $\bar{n}=0$ for a maximum $\mathcal{J}$. For a given squeezing $r$, the parameter $k$ is bounded by $|k|\leq 2 \tanh 2r$, where the bounds are saturated at $\theta-\phi= \pm \frac{\pi}{4}$, 
at which the rescaling in Eq. (13) reads $( x^{\prime}, y^{\prime} ) =\frac{1}{\kappa}(x,y)$ with $\kappa=\sqrt{\frac{1+2\bar{n}}{2\cosh 2r}}$ (the same $\kappa$ below Eq. (5) in the main Letter).

We may be restricted only to $k>0$, as the maximum for $k<0$ is equivalentely identified by $y'\rightarrow-y'$ in Eq. (14). 
Furthermore, as we show below, an optimal $\mathcal{J}$ monotonically increases with $k$. This means that for a given squeezed state an optimal value occurs at 
\begin{eqnarray}
\theta-\phi= \frac{\pi}{4},
\end{eqnarray}
i.e., the rectangle is rotated $45^{\rm o}$ from the squeezed axis. Intuitively, this choice of rectangle makes a sense because our test of nonclassicality relies on the non-factorizability of Wigner function. (See the paragraphs below Eq. (1) in the main Letter.) For instance, if we take $\theta-\phi=0$, the Wigner function in Eq. (14) is factorized to $W_{\sigma} ( x^{\prime}, y^{\prime} )=W( x^{\prime}) W ( y^{\prime} ),$
leading to no violation of Eq. (3) in the main Letter.

---{\bf Condition for an optimal $\mathcal{J}$}\\  
Now the problem is reduced to the optimization of 
	\begin{align}
		\mathcal{F}_{k} = & e^{-x_{0}^{2}-y_{0}^{2}+kx_{0}y_{0}} + e^{-x_{0}^{2}-y_{1}^{2}+kx_{0}y_{1}} \nonumber \\
		+ & e^{-x_{1}^{2}-y_{0}^{2}+kx_{1}y_{0}} - e^{-x_{1}^{2}-y_{1}^{2}+kx_{1}y_{1}},
	\end{align}
where $0 \leq k \leq 2$. The minimum $\mathcal{F}_{k}$ must be trivially $-1$ for all Gaussian states:
$\mathcal{F}_{k}$ consists of four exponential terms, each taking a value in [0,1] with $x^2+y^2-kxy=(x-\frac{ky}{2})^2+(1-\frac{k^2}{4})y^2\ge0$. The minimum can occur at, e.g., $x_0=y_0\rightarrow\infty$ and $x_1=y_1=0$, which makes the first three terms zero and the last $-1$ in $\mathcal{F}_{k}$.

For a maximum, we obtain the partial derivatives
	\begin{align}
		\frac{\partial \mathcal{F}_{k}}{\partial x_{0}} & = e^{- x_{0}^{2} - y_{0}^{2} + k x_{0} y_{0}} ( - 2 x_{0} + k y_{0} ) \nonumber \\
		& + e^{- x_{0}^{2} - y_{1}^{2} + k x_{0} y_{1}} ( - 2 x_{0} + k y_{1} ), \nonumber \\
		\frac{\partial \mathcal{F}_{k}}{\partial y_{0}} & = e^{- x_{0}^{2} - y_{0}^{2} + k x_{0} y_{0}} ( k x_{0} - 2 y_{0} ) \nonumber \\
		& + e^{- x_{1}^{2} - y_{0}^{2} + k x_{1} y_{0}} ( k x_{1} - 2 y_{0} ), \nonumber \\
		\frac{\partial \mathcal{F}_{k}}{\partial x_{1}} & = e^{- x_{1}^{2} - y_{0}^{2} + k x_{1} y_{0}} ( - 2 x_{1} + k y_{0} ) \nonumber \\
		& - e^{- x_{1}^{2} - y_{1}^{2} + k x_{1} y_{1}} ( - 2 x_{1} + k y_{1} ), \nonumber \\
		\frac{\partial \mathcal{F}_{k}}{\partial y_{1}} & = e^{- x_{0}^{2} - y_{1}^{2} + k x_{0} y_{1}} ( k x_{0} - 2 y_{1} ) \nonumber \\
		& - e^{- x_{1}^{2} - y_{1}^{2} + k x_{1} y_{1}} ( k x_{1} - 2 y_{1} ).
	\end{align}
Using the conditions $\frac{\partial \mathcal{F}_{k}}{\partial x_{0}} = \frac{\partial \mathcal{F}_{k}}{\partial y_{0}} = \frac{\partial \mathcal{F}_{k}}{\partial x_{1}} = \frac{\partial \mathcal{F}_{k}}{\partial y_{1}}$ = 0 at optimal points, we have 
	\begin{align} \label{eq:extrema1}
		& \frac{\partial \mathcal{F}_{k}}{\partial x_{0}} (kx_{0}-2y_{0}) (-2x_{1}+ky_{0}) (kx_{1}-2y_{1}) \nonumber \\
		- & \frac{\partial \mathcal{F}_{k}}{\partial y_{0}} (-2x_{0}+ky_{0}) (-2x_{1}+ky_{0}) (kx_{1}-2y_{1}) \nonumber \\
		+ & \frac{\partial \mathcal{F}_{k}}{\partial x_{1}} (-2x_{0}+ky_{0}) (kx_{1}-2y_{0}) (kx_{1}-2y_{1}) \nonumber \\
		- & \frac{\partial \mathcal{F}_{k}}{\partial y_{1}} (-2x_{0}+ky_{0}) (kx_{1}-2y_{0}) (-2x_{1}+ky_{1}) \nonumber \\
		= & 2k(4-k^{2}) (x_{0}-x_{1}) (y_{0}-y_{1}) (x_{0}x_{1}-y_{0}y_{1}) e^{-x_{0}^{2}+kx_{0}y_{1}-y_{1}^{2}} \nonumber \\
		= & 0.
	\end{align}
As $x_{0} = x_{1}$ and $y_{0} = y_{1}$ give only a classical value, $\mathcal{F}_{k} = 2 e^{-x_{0}^{2}-y_{0}^{2}+kx_{0}y_{0}} \leq 2$, the optimal points we seek must satisfy
$x_{0} x_{1} = y_{0} y_{1}$, which is equivalent to 
\begin{eqnarray}
\frac{x_{0}}{y_{0}}=\frac{y_{1}}{x_{1}}\equiv t,
\end{eqnarray}
thus $x_{0}=t{y_{0}}$ and $y_{1}=t{x_{1}}$. Furthermore, we may set $x_{1}=cy_{0}$, which enables us to cover all pairs of $\{x_{1},y_{0}\}$ by scanning $c$ in the entire range of real values.

Then, setting $( x_{0}, y_{0}, x_{1}, y_{1} ) = ( ty, y, cy, tcy )$, $\mathcal{F}_{k}$ in Eq. (17) reads
\begin{eqnarray}
\mathcal{F}_{k}(y,c,t)=&&e^{-(t^2+1-kt)y^2}-e^{-c^2(t^2+1-kt)y^2}\nonumber\\&&+e^{-t^2(c^2+1-kc)y^2}+e^{-(c^2+1-kc)y^2}.
\end{eqnarray}  
By defining $y'\equiv y\sqrt{t^2+1-kt}$, 
the above becomes
\begin{eqnarray}
\mathcal{{\tilde F}}_{k}(y',c,t)=e^{-y'^2}-e^{-c^2y'^2}+e^{-t^2f(c,t,k)y'^2}+e^{-f(c,t,k)y'^2},\nonumber\\
\end{eqnarray}  
where $f(c,t,k)\equiv\frac{c^2+1-kc}{t^2+1-kt}$. Obviously, if one finds a global maximum for $\mathcal{{\tilde F}}$ in Eq. (22) with a set of values $\{y',c,t,k\}$, the same maximum for $\mathcal{F}$ in Eq. (21) must occur at $\{y=\frac{y'}{\sqrt{t^2+1-kt}},c,t,k\}$. Now, taking a partial derivative of $\mathcal{{\tilde F}}$ over $k$, 
\begin{eqnarray}
\frac{\partial\mathcal{{\tilde F}}}{\partial k}=-2y'\frac{(ct-1)(c-t)}{(t^2+1-kt)^2}\left(t^2e^{-t^2f(c,t,k)y'^2}+e^{-f(c,t,k)y'^2}\right),\nonumber\\
\end{eqnarray}
we see that $\mathcal{{\tilde F}}_{k}$ is a monotonic function of $k\in[0,2]$ for any given values of $\{y',c,t\}$. This means that we only need to consider either $k=0$ or $k=2$ at each point of parameter space to find a global maximum. That is, $\max\{\mathcal{{\tilde F}}_{k=0}(y',c,t),\mathcal{{\tilde F}}_{k=2}(y',c,t)\}$ becomes the global maximum of $\mathcal{{\tilde F}}$ by covering the whole ranges of $\{y',c,t\}$. As $k=0$ is the case of coherent state yielding only a classical bound 2, we conclude that the case of $k=2$, i.e. an infinitely squeezed state, gives the maximum bound for all Gaussian states.

	\begin{figure}[!t]
		\includegraphics[scale=0.4]{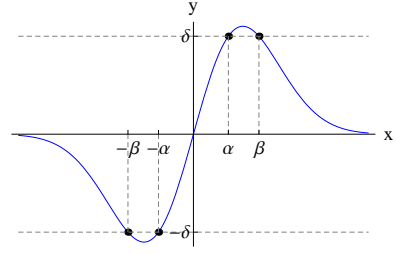}
		\caption{Plot for $y = x e^{- x^{2}}$. The number of solutions for $x e^{- x^{2}} = \delta$ is two at most.}
		\label{fig:F2}
	\end{figure}

---{\bf Maximum Gaussian bound $\frac{8}{3^{9/8}}$}\\
Coming back to Eq. (18) with $k=2$, the partial derivatives are rather simplified to
	\begin{align}
		\frac{\partial \mathcal{F}_{2}}{\partial x_{0}} & = - 2 e^{- ( x_{0} - y_{0} )^{2}} ( x_{0} - y_{0} ) - 2 e^{- ( x_{0} - y_{1} )^{2}} ( x_{0} - y_{1} ), \nonumber \\
		\frac{\partial \mathcal{F}_{2}}{\partial y_{0}} & = 2 e^{- ( x_{0} - y_{0} )^{2}} ( x_{0} - y_{0} ) + 2 e^{- ( x_{1} - y_{0} )^{2}} ( x_{1} - y_{0} ), \nonumber \\
		\frac{\partial \mathcal{F}_{2}}{\partial x_{1}} & = - 2 e^{- ( x_{1} - y_{0} )^{2}} ( x_{1} - y_{0} ) + 2 e^{- ( x_{1} - y_{1} )^{2}} ( x_{1} - y_{1} ), \nonumber \\
		\frac{\partial \mathcal{F}_{2}}{\partial y_{1}} & = 2 e^{- ( x_{0} - y_{1} )^{2}} ( x_{0} - y_{1} ) - 2 e^{- ( x_{1} - y_{1} )^{2}} ( x_{1} - y_{1} ).	
	\end{align}
A trivial solution for $\frac{\partial \mathcal{F}_{2}}{\partial x_{0}}  = \frac{\partial \mathcal{F}_{2}}{\partial y_{0}} = \frac{\partial \mathcal{F}_{2}}{\partial x_{1}} = \frac{\partial \mathcal{F}_{2}}{\partial y_{1}} = 0$ is $x_{0} = y_{0} = x_{1} = y_{1}$ which only yields $\mathcal{F}_{2} = 2$. To find the other solution, we make two observations. (1) The above equations impose an ordering for optimal parameters.  If we have $x_{1} > y_{1}$ or $x_{0} < y_{0}$, substituting the condition to the equations yield $x_{1} > y_{0} > x_{0} > y_{1}$ or $x_{1} < y_{0} < x_{0} < y_{1}$, respectively. From now on, we assume $x_{1} > y_{1}$ without loss of generality. (2) Every equation is in the same form: $\alpha e^{- \alpha^{2}} - \beta e^{- \beta^{2}} = 0$. From $\frac{\partial \mathcal{F}_{2}}{\partial x_{1}} = \frac{\partial \mathcal{F}_{2}}{\partial y_{1}} = 0$, we obtain $x_{0} - y_{1} = x_{1} - y_{0}$ in view of the ordering of optimal parameters and Fig.~\ref{fig:F2}. As can be seen from Fig.~\ref{fig:F2}, the number of solutions for $x e^{- x^{2}} = \delta$ is at most two. In addition, none of $x_{0} - y_{1}$ and $x_{1} - y_{0}$ can be identical to $x_{1} - y_{1}$ because of the ordering. The only possible case is then $x_{0} - y_{1} = x_{1} - y_{0} = \alpha$ and $x_{1} - y_{1} = \beta$ where $\alpha < \beta$ with $\alpha e^{- \alpha^{2}} = \beta e^{- \beta^{2}}$. Using a similiar argument to $\frac{\partial \mathcal{F}}{\partial x_{0}} = \frac{\partial \mathcal{F}}{\partial y_{0}} = 0$, we also have $x_{0} - y_{0} = - \alpha$. Combining the results, we have $\beta = 3 \alpha$; $x_{1} - y_{1} = ( x_{0} - y_{1} ) + ( x_{1} - y_{0} ) - ( x_{0} - y_{0} )$. It enables us to rewrite the function as $\mathcal{F}_{2} = 3 e^{- \alpha^{2}} - e^{- 9 \alpha^{2}}$, which gives the maximum $\frac{8}{3^{9/8}}$ at $\alpha = \sqrt{\frac{\log 3}{8}}$. Combined with the condition $x_0x_1=y_0y_1$, it also gives the optimal points $x_0=-y_0=-\frac{1}{2}\alpha$ and $x_1=-y_1=\frac{3}{2}\alpha$. 


---{\bf Optimal $\mathcal{J}$ for finitely squeezed states}\\
Let us now find an optimal $\mathcal{F}_{k}$ in Eq. (21) for each $k\in[0,2]$.
 Its partial derivatives are given by
	\begin{align}
		\frac{\partial \mathcal{F}_{k}}{\partial t} & = -y^{2} \{ (2t-k) e^{-(t^{2}-kt+1)y^{2}} \nonumber \\
		& + 2t(c^{2}-kc+1) e^{-t^{2}(c^{2}-kc+1)y^{2}} \nonumber \\
		& - (2t-k) c^{2} e^{-(t^{2}-kt+1)c^{2}y^{2}} \}, \nonumber \\
		\frac{\partial \mathcal{F}_{k}}{\partial c} & = -y^{2} \{ (2c-k) e^{-(c^{2}-kc+1)y^{2}} \nonumber \\
		& + t^{2} (2c-k) e^{-t^{2}(c^{2}-kc+1)y^{2}} \nonumber \\
		& - 2c(t^{2}-kt+1) e^{-(t^{2}-kt+1)c^{2}y^{2}} \}, \nonumber \\
		\frac{\partial \mathcal{F}_{k}}{\partial y} & = -2y \{ (t^{2}-kt+1)  e^{-(t^{2}-kt+1)y^{2}} \nonumber \\
		& + (c^{2}-kc+1) e^{-(c^{2}-kc+1)y^{2}} + \nonumber \\
		& + t^{2}(c^{2}-kc+1) e^{-t^{2}(c^{2}-kc+1)y^{2}} \nonumber \\
		& - (t^{2}-kt+1)c^{2} e^{-(t^{2}-kt+1)c^{2}y^{2}} ) \}.
	\end{align}
In the case of $k=2$ considered before, the optimal points were given as $x_0=-y_0=-\frac{1}{2}\alpha$ and $x_1=-y_1=\frac{3}{2}\alpha$. This indicates $t=-1$, recalling the parametrization $( x_{0}, y_{0}, x_{1}, y_{1} ) = ( ty, y, cy, tcy )$. Indeed, we find that a set of optimal points  
satisfy $\frac{\partial \mathcal{F}_{k}}{\partial t} = \frac{\partial \mathcal{F}_{k}}{\partial c} = \frac{\partial \mathcal{F}_{k}}{\partial x} = 0$, with $t=-1$ and 
	\begin{align}
		e^{-\{(1+k)c^{2}+kc-1\}y^{2}} & = \frac{2c-k}{c(k+2)}, \nonumber \\
		e^{-(1+c)(1-c+k)y^{2}} & = \frac{kc-2}{k+2},
	\end{align}
which gives
	\begin{align}
		& \{(1+k)c^{2}+kc-1\} \log \frac{kc-2}{k+2} \nonumber \\
		& = (1+c)(1-c+k) \log \frac{2c-k}{c(k+2)}.
	\end{align}
For a given $k\in[0,2]$, we can numerically obtain $c$ from Eq. (27), and then $y$ from Eq. (26). (The parameters $z_i$ introduced below Eq. (5) in the main Letter then reads $z_0=y$ and $z_1=cy$.) The resulting maximum $\mathcal{F}_{k}$ (blue solid curve) is shown in Fig.~2 (a) of the main Letter, which is also in agreement with the result obtained independently from a full numerical optimaization (red dots) by scanning all real ranges of $\{t,c,y\}$ in Eq. (21). 

As an analytical example, Eq. (27) gives $c=3$ for $k=2$, which recovers the optimal points $x_0=-y_0=-\frac{1}{2}\alpha$ and $x_1=-y_1=\frac{3}{2}\alpha$ obtained independently in the previous subsection. For a mixed state with $\mu<1$, we only need to multiply a factor $\mu=\frac{1}{1+2\bar{n}}$, as seen from Eq. (14), to a result for a pure state with the same $r$.

\section{B. Right triangle test}
We here introduce another nonclassicality test taking {\it three} phase-space points into consideration, which has no analogue in nonlocality test, as
	\begin{equation}
		\mathcal{J}^{\prime} [ \rho ] \equiv \frac{\pi}{2} \{ W_{\rho}^{\theta} ( x_{1}, y_{0} ) + W_{\rho}^{\theta} ( x_{0}, y_{1} ) - W_{\rho}^{\theta} ( x_{1}, y_{1} ) \},
	\end{equation}
where the three points may form a triangle rotated by an angle $\theta$ from $q$-axis in phase space. Our test is linear with respect to a convex mixture of quantum states, i.e. $\mathcal{J}^{\prime} [ \sum_{j} p_{j} \rho_{j} ] = \sum_{j} p_{j} \mathcal{J}^{\prime} [ \rho_{j} ]$, and displacement does not change the optimal value of $\mathcal{J}^{\prime}$, like the four-points test in the previous section. Investigating a vacuum state thus suffices to find its classical bound as
	\begin{equation}
		\mathcal{J}^{\prime} [ \ketbra{0}{0} ]  =  e^{- 2 x_{1}^{2}} e^{- 2 y_{0}^{2}} + ( e^{- 2 x_{0}^{2}} - e^{- 2 x_{1}^{2}} ) e^{- 2 y_{1}^{2}}.
	\end{equation}
Regardless of the sign of $e^{- 2 x_{0}^{2}} - e^{- 2 x_{1}^{2}}$, we find the classical bound as $\mathcal{J}^{\prime} [ \rho_{\mathrm{cl}} ] \leq 1$; 

(i) if $e^{- 2 x_{0}^{2}} - e^{- 2 x_{1}^{2}} \leq 0$, we have $\mathcal{J}^{\prime} [ \ketbra{0}{0} ] \leq e^{- 2 x_{1}^{2}} e^{- 2 y_{0}^{2}} \leq 1$. 

(ii) if $e^{- 2 x_{0}^{2}} - e^{- 2 x_{1}^{2}} > 0$, we have
	\begin{align}
		\mathcal{J}^{\prime} [ \ketbra{0}{0} ] & \leq e^{- 2 x_{1}^{2}} e^{- 2 y_{0}^{2}} + e^{- 2 x_{0}^{2}} - e^{- 2 x_{1}^{2}} \nonumber \\
		& = - e^{- 2 x_{1}^{2}} ( 1 - e^{- 2 y_{0}^{2}} ) + e^{- 2 x_{0}^{2}} \leq 1.
	\end{align}
	
In contrast, there does not exist a useful nonlocality test constructed out of a linear sum of three product values only, as $B'=a_1b_0+a_0b_1-a_1b_1$. One can readily find a set of local values that achieve a maximum value 3 ($a_0=1,a_1=-1$ and $b_0=-1,b_1=1$) and a minimum value -3 ($a_0=-1,a_1=1$ and $b_0=-1,b_1=1$), respectively. Therefore, our proposed test here taking only three points is a unique nonclassicality test having no analogue in nonlocality test.

For a given Gaussian state with $( \alpha, r, \phi, \bar{n} )$, the maximal value for this triangle test occurs at,
	\begin{align} \label{eq:op2}
		x_{j} & = \alpha_{x} + (-1)^{j} z_{j}^{\prime} \kappa \nonumber \\
		y_{j} & = \alpha_{y} - (-1)^{j} z_{j}^{\prime} \kappa,
	\end{align}
where $j \in \{ 0, 1 \}$, $\theta = \phi + \frac{\pi}{4}$, $\kappa = \sqrt{\frac{2 \bar{n} + 1}{2 \cosh 2r}}$, and $z_j$ can be found analytically, similar to four-points test. 

---{\bf Optimal $\mathcal{J'}$}\\
Here, after the rescaling of coordinates 
$\frac{1}{\kappa}(x,y)\rightarrow(x,y)$, we need to optimize 
	\begin{equation}
		\mathcal{H}_{k} = e^{-x_{0}^{2}-y_{1}^{2}+kx_{0}y_{1}} + e^{-x_{1}^{2}-y_{0}^{2}+kx_{1}y_{0}} - e^{-x_{1}^{2}-y_{1}^{2}+kx_{1}y_{1}}.
	\end{equation}
The conditions $\frac{\partial \mathcal{H}_{k}}{\partial x_{0}}=\frac{\partial \mathcal{H}_{k}}{\partial y_{0}}=\frac{\partial \mathcal{H}_{k}}{\partial x_{1}}=\frac{\partial \mathcal{H}_{k}}{\partial y_{1}}=0$	
immediately give $x_{1} = \frac{2}{k} y_{0}$ and $y_{1} = \frac{2}{k} x_{0}$. Then, using a parametrization $( x_{0}, y_{0}, x_{1}, y_{1} ) = ( ty, y, \frac{2}{k} y, \frac{2}{k} ty )$, we consider $\mathcal{H}_{k}(y,t) =  e^{-(\frac{4}{k^{2}}-1)y^{2}} + e^{-t^{2}(\frac{4}{k^{2}}-1)y^{2}} - e^{-(t^{2}-kt+1)\frac{4}{k^{2}}y^{2}}$, which leads to $\frac{\partial \mathcal{H}_{k}}{\partial t} = \frac{\partial \mathcal{H}_{k}}{\partial y} = 0$ as
	\begin{align}
		( 1 - \frac{k^{2}}{4} ) \frac{2t}{2t-k} & = \exp [ - ( t - \frac{2}{k} )^{2} y^{2} ], \nonumber \\
		( 1 - \frac{k^{2}}{4} ) \frac{2}{2-kt} & = \exp [ - ( \frac{2t}{k} - 1 )^{2} y^{2} ].
	\end{align}
The combination of the above two equations gives
	\begin{align}
		( kt - 2 )^{2} \log [ ( 1 - \frac{k^{2}}{4} ) \frac{2}{2-kt} ] = ( 2t - k )^{2} \log [ ( 1 - \frac{k^{2}}{4} ) \frac{2t}{2t-k} ].
	\end{align}

Noting that $\mathcal{H}_{k}(y,-|t|)\ge\mathcal{H}_{k}(y,|t|)$, an optimal point must occur for $t<0$. We have numerically confirmed that $t=-1$ is the only negative solution to Eq. (34) for any $k\in[0,2]$. Then, we obtain $y^{2} = \frac{1}{(1+\frac{2}{k})^{2}} \log [ \frac{1}{1 - \frac{k}{2}} ]$ and the maximum $\mathcal{H}_{k} = \frac{1}{2^{\frac{4}{2+k}}} (2-k)^{\frac{2-k}{2+k}} (2+k)$ monotonically increasing to 2 with $k$. The optimizing parameters in Eq. (31) read $z_0'=y$ and $z_1'=\frac{2}{k}y$.



Similar to the rectangle test using four phase-space points, the triangle test detects every pure Gaussian nonclassical state. Moreover, Fig.~2 (b) of the main Letter clearly demonstrates that the triangle test outperforms the rectangle test to detect nonclassical Gaussian states. 

Using the fact that no Gaussian state overcomes the bound $\mathcal{B}_{G}^{\prime} = 2$, we can also construct a genuine quantum non-Gaussianity test as
	\begin{equation}
		\mathcal{W}^{\prime} \equiv  \mathcal{J}^{\prime} [ \hat{U}_{G} \rho \hat{U}_{G}^{\dag} ] - 2,
	\end{equation}
where a positive value of the witness, $\mathcal{W}^{\prime} > 0$ for a certain $\hat{U}_{G}$, manifests that the state is genuinely quantum non-Gaussian. In order to beat this bound, $W_{\hat{U}_{G} \rho \hat{U}_{G}^{\dag}} ( q, p )$ must necessarily take a negative value. While this test may be interesting in itself, it would not be practically significant compared with the rectangle or the parallelogram test in the main Letter as the latters can detect quantum non-Gaussianity of a positive Wigner function.

\section{C. Violation of lower bound $-1$ for a nonclassicality test}
\begin{figure}[!t]
		\includegraphics[scale=0.5]{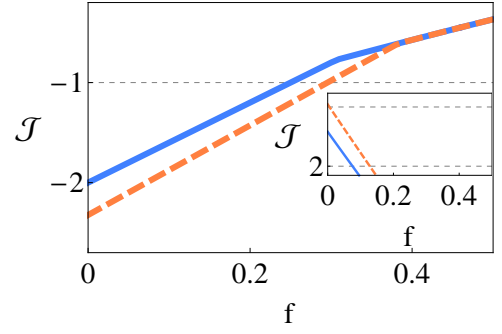}
		\caption{A noisy single photon state, $f \ketbra{0}{0} + (1-f) \ketbra{1}{1}$ under rectangle (blue solid) and parallelogram (red dashed) tests. The violation of lower bounds demonstrates nonclassicality and non-Gaussianity better than that of upper bounds (inset). That is, the lower bounds detect the state in the ranges $0 \leq f < 0.25$  and $0 \leq f \lesssim 0.33$ under rectangle and the parallelogram tests, respectively, while the upper bounds do so in $0 \leq f \lesssim 0.08$ and $0 \leq f < 0.01$.}
		\label{fig:F4}
	\end{figure}

The lower bound $-1$ of our nonclassicality test, Eq. (3) in the main Letter, can be more useful than the upper bound 2 if there exists a substantial negativity in the Wigner function of the original state, even when it is mixed with some noise. For example, we plot the results for a noisy single photon state, $f \ketbra{0}{0} + (1-f) \ketbra{1}{1}$, in Fig.~\ref{fig:F4}. It shows that the violation of the lower bound occurs for a broader range of $f$ than the violation of the upper bound. 


\section{D. Maximal value of $\mathcal{J} [ \rho ]$ in Eq. (2) by quantum non-Gaussian states}
In order to find out a maximum possible value of $\mathcal{J}$ by a quantum non-Gaussian state in Eq.~(2), we here solve an eigenvalue problem $\mathcal{H} \ket{\psi} = \lambda \ket{\psi}$, where the Hermitian operator $\mathcal{H}$ is defined as
	\begin{equation} \label{eq:H}
		\mathcal{H}\equiv \sum_{a = 0}^{1} \sum_{b = 0}^{1} (-1)^{ab} \hat{D} ( q_{a} + i p_{b} ) (-1)^{\hat{n}} \hat{D}^{\dag} ( q_{a} + i p_{b} ). 
	\end{equation}
We note that the four terms in $\mathcal{H}$ contributing to a linear sum have a different structure from	
those in the Bell operator to test nonlocality,  $\hat{B}=\hat{a}_0\hat{b}_0+\hat{a}_0\hat{b}_1+\hat{a}_1\hat{b}_0-\hat{a}_1\hat{b}_1$. 
In the latter case, using $\hat{a}_i^2=\hat{b}_j^2=I$ ($i,j=0,1$) and the commutativity of	{\it spatially separated} observables $[\hat{a}_i,\hat{b}_j]=0$, we are led to 
$\hat{B}^2=4I+[\hat{a}_0,\hat{a}_1]\otimes[\hat{b}_1,\hat{b}_0]$, from which the maximum $\langle\hat{B}^2\rangle_{\rm max}=8$ known as the Cirelson bound \cite{Cirelson1980} can be derived. In contrast, each term of $\mathcal{H}$ in our test is not factorized to a product of two commuting observables (unlike the form $\hat{a}_i\hat{b}_j$ in $\hat{B}$), which instead involves {\it local} non-commuting observables $\hat{q},\hat{p}$ and $\hat{n}$. This indeed makes it possible to achieve a maximum eigenvalue of $\mathcal{H}$ above $2\sqrt{2}$, as we show below.

	\begin{figure}[!]
		\includegraphics[scale=0.6]{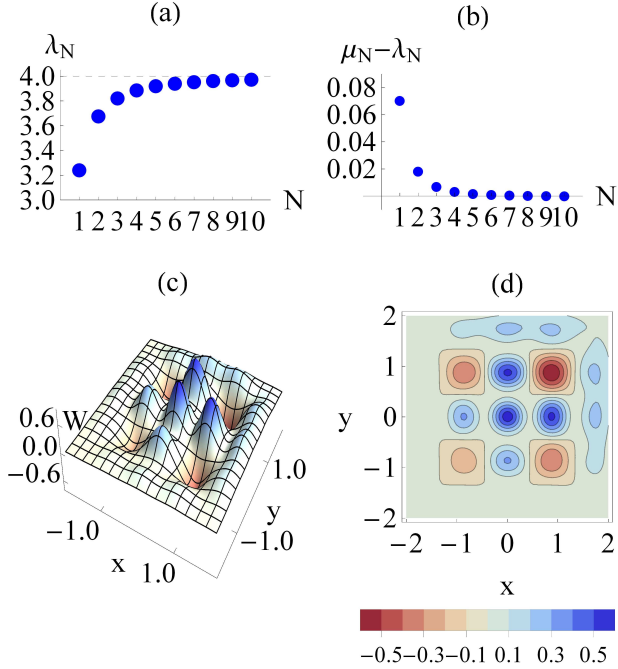}
		\caption{(a) Maximum eigenvalue $\lambda_N$ of $\mathcal{H}$ in Eq.~(36) that is numerically obtained by truncating $n$ and $m$ up to $N$ in the recurrence relations of Eq. (38). (b) Expectation value $\mu_N=\langle\mathcal{H}\rangle$ of a truncated superposition of $(2N+1) \times (2N+1)$ coherent states in Eq.~(\ref{eq:L1}) compared with $\lambda_N$.  (c) Wigner function and (d) its contour plot for a superposition of $3\times3$ coherent states with a lattice size $2d = \sqrt{\pi}$ in Eq.~(\ref{eq:L1}).}
		\label{fig:EV}
	\end{figure}

We may set $( q_{0}, p_{0}, q_{1}, p_{1} )$ to be $( 0, 0, d, d )$ without loss of generality. Note that we can always transform arbitrary four variables $( q_{0}, p_{0}, q_{1}, p_{1} )$ to $( 0, 0, d, d )$ by Gaussian unitary operations, which do not affect the maximum value of $\mathcal{J}$ achieved by a non-Gaussian state. We seek a solution in a form
	\begin{equation} \label{eq:ES1}
		\ket{\psi} = \sum_{n = - \infty}^{\infty} \sum_{m = - \infty}^{\infty} C_{n,m} \ket{2d(n+im)},
	\end{equation}
where $\ket{2d(n+im)}$ is a coherent state ($n$ and $m$: integers), that is, a superposition of coherent states in a 2-dimensional lattice of size $2d$. 
Using Eqs.~\eqref{eq:H} and \eqref{eq:ES1} and the identity $\hat{D} ( \alpha ) (-1)^{\hat{n}} \hat{D}^{\dag} ( \alpha ) \ket{\gamma} = e^{- \alpha \gamma^{*} + \alpha^{*} \gamma} \ket{2 \alpha - \gamma}$, we obtain a recurrence relation
	\begin{align}
		\lambda C_{n,m} = & C_{-n,-m} + e^{-4id^{2}m} C_{-n+1,-m} + e^{4id^{2}n} C_{-n,-m+1} \nonumber \\
		& - e^{-4id^{2}(m-n)} C_{-n+1,-m+1}.
	\end{align}
In Fig.~\ref{fig:EV} (a), we show the maximum eigenvalues numerically obtained by a truncation of numbers up to $N$ in the recurrence relations, which turns out to approach 4 with $N$ increasing for a choice of $d^{2} = \frac{R \pi}{2} + \frac{\pi}{4}$ ($R$: integer). To check whether these values can actually be achieved, we compare $\lambda_{N}$ and $\mu_{N} \equiv \bra{\psi^{N}} \mathcal{H} \ket{\psi^{N}} / \braket{\psi^{N}}{\psi^{N}}$ where
	\begin{equation} \label{eq:L1}
		\ket{\psi^{N}} = \sum_{n=-N}^{N} \sum_{m=-N}^{N} C_{n,m} \ket{2d(n+im)},
	\end{equation}
which is shown in Fig.~\ref{fig:EV} (b).

As an illustration, we show the case of $N=1$, i.e. a superposition of $3\times3$ coherent states in Fig.~\ref{fig:EV} (c) and (d), which achieves a value of $\mathcal{J} [ \rho ]\approx3.32$.

\section{E. Experimental realization} 
\begin{figure}
	\includegraphics[scale=0.58]{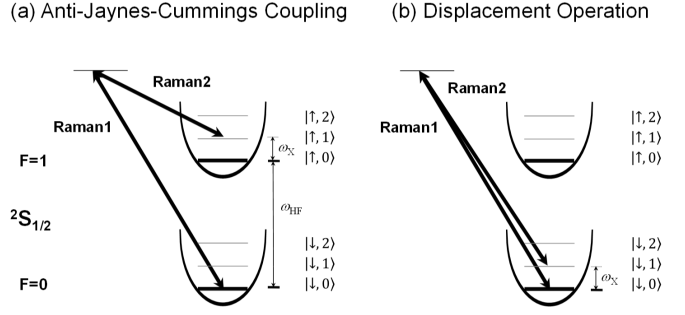}
		\caption{ Experimental configuration with external driving fields to produce (a) the anti-Jaynes-Cummings interaction between internal and motional degrees of freedom and (b) the displacement operation on the motional mode. }
		\label{fig:setup}
	\end{figure}

The control and manipulation of phonon mode can be achieved by the anti-Jaynes-Cumming interaction that couples the phonon mode $\hat{a}$ to the internal states represented by Pauli operators $\hat\sigma_i$, which is given by $H_{\rm aJC} = \frac{\eta \Omega}{2} \hat{a}^{\dagger} \hat{\sigma}_{+} + {\rm h.c.}$. It can be realized by two counter-propagating Raman laser beams that have the beat frequency near $\omega_{\rm HF} + \omega_{\rm X}$ as shown in Fig. \ref{fig:setup} (a). On the other hand, for a displacement operation on the phonon mode, the Hamiltonian $H_{D} = \frac{\eta\Omega_{D} e^{i \phi}}{2} \hat{a}^{\dagger} + {\rm h.c.}$ can be realized in the same experimental configuration, with the beat frequency near $\omega_{\rm X}$ and the phase $\phi$, as shown in Fig. \ref{fig:setup} (b) \cite{KKim2014}.

\begin{figure}
	\includegraphics[scale=0.25]{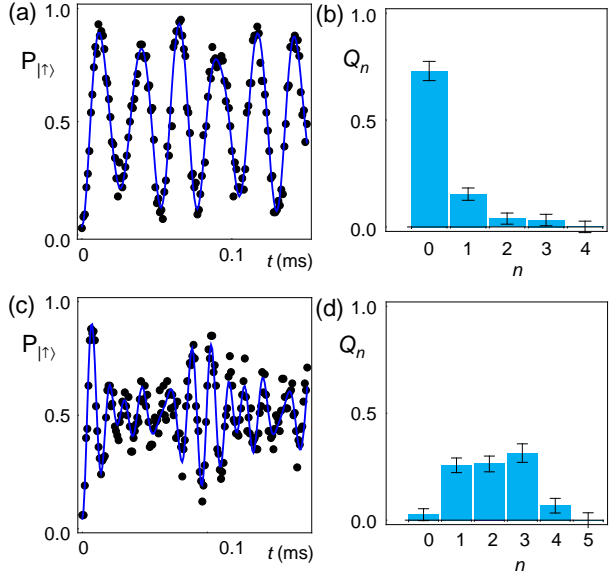}
		\caption{ (a)(c) Typical time evolutions with the application of $H_{\rm aJC}$ and (b)(d) phonon distributions after fitting the signal to Eq. (7) in the main Letter. (a)(b) are the results after displacement operation on a vacuum state $\ket{n=0}$ to (0.331,0.331) in phase space, while (c)(d) are the ones after the same displacement operation on the state $\ket{n=2}$. }
		\label{fig:typicalsig}
	\end{figure}

For the measurement of $\mathcal{J}$ and Wigner function, first we perform a displacement operation on an initial state, then apply the anti-Jaynes-Cumming interaction to obtain the phonon number distributions, which provide the value of the parity, $\sum_{n} (-1)^n Q_{n}$. Figure \ref{fig:typicalsig} shows typical time evolutions of $P_{|\uparrow\rangle}$, i.e., population in the state $|\uparrow\rangle$ under the operation of anti-Jaynes-Cummings interaction $H_{\rm aJC}$ and the phonon-number distributions $Q_{n}$ that are obtained by fitting the time evolution signal to Eq. (7) in the main Letter.

We also experimentally measure the density matrices of the state $\ketbra{0}{0}$ and $\ketbra{2}{2}$ and construct their Wigner functions using the density matrices. To measure the density matrix, we basically follow the method in Ref. \cite{WWineland96} and apply the iterative maximum-likelihood method \cite{Jezek01} to avoid unphysical, non-positive, matrix by the direct reconstruction. In experiment, we displace the state $\rho$ to eight different angles with the fixed amplitude of $\alpha=0.8$.

	\begin{figure}
	\includegraphics[scale=0.45]{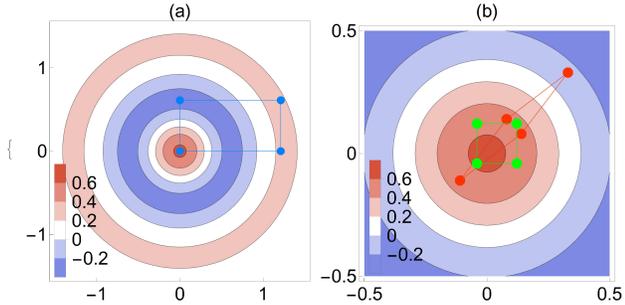}
		\caption{ Contour plot of Wigner function for a Fock state $|2\rangle$, with (a) four points chosen for a nonclassicality test, and (b) four points (green) of rectangle transformed to those (red) of parallelogram. Red points were used in our experiment of non-Gaussianity test. }
		\label{fig:contour}
	\end{figure}		
---{\bf Choice of phase-space points for test}

For our test of non-Gaussian states, there would not be a unique, systematic, way to choose optimal phase-space points, similar to the situation with nonlocality tests. However, we may give a rough description of choosing appropriate points as follows. For a rectangle test, as three terms contribute positively and one term negatively in $\mathcal{J}$, we may configure the rectangle such that one vertex corresponds to a negative value of Wigner function. An example of Fock state $\ket{2}$ is illustrated in Fig. \ref{fig:contour} (a), together with the choice of four phase-space points. The left-upper point is placed at a negative Wigner function. 
	
On the other hand, when a pure non-Gaussian state is mixed with noise, e.g. $f \ketbra{0}{0} + (1-f) \ketbra{2}{2}$ in our experiment, the parallelogram test provides a much more efficient method. We may first choose a rectangle with a modest size (green dots in Fig. \ref{fig:contour} (b) ) and then shift the points (squeeze the rectangle) according to the rule given below Eq. (6) in the main Letter, resulting in a parallelogram (red dots in Fig.\ref{fig:contour} (b)). Note that one red point now reaches a negative region of $\ket{2}$ by this transformation, although it is degraded by the vacuum state $\ket{0}$ (not shown) that has a positive-definite Wigner function.
Of course, we need optimization over the initial rectangles followed by squeezing, but it can be done at least numerically with some prior information on the produced state. This is how we chose the phase-space points for $f \ketbra{0}{0} + (1-f) \ketbra{2}{2}$. Even when selected points are not strictly optimal, we can experimentally demonstrate the genuine non-Gaussianity to a large extent, as the inequality (6) of the main Letter is satisfied in a rather broad parameter region. In fact, this was the case with our experiment, taking four (non-optimal) phase-space points well-separated for an easy control of displacement.


\begin{thebibliography}{99}


\bibitem{Wigner1932} E. Wigner, Phys. Rev. \textbf{40}, 749 (1932).

\bibitem{Cerf} Quantum Information with Continuous Variables of Atoms and Light, edited by N. Cerf, G. Leuchs, and E. S. Polzik (Imperial College Press, London, 2007).

\bibitem{Zurek2003} W. H. Zurek, Rev. Mod. Phys. \textbf{75}, 715 (2003).

\bibitem{Barnett} S. M. Barnett and P. M. Radmore, {\it Methods in Theoretical Quantum Optics}, Oxford University Press (2003).

\bibitem{Braunstein2005} S. L. Braunstein and P. Van Loock, Rev. Mod. Phys. \textbf{77}, 513 (2005).

\bibitem{Lvovsky2009} A. I. Lvovsky and M. G. Raymer, Rev. Mod. Phys. \textbf{81}, 299 (2009).
\bibitem{Richter} Th. Richter and W. Vogel \prl {\bf 89}, 283601 (2002).
\bibitem{Mari} A. Mari, K. Kieling, B. Melholt Nielsen, E. S. Polzik, and J. Eisert, \prl {\bf 106}, 010403 (2011).


\bibitem{Bell} J. S. Bell, {\it Speakable and Unspeakable in Quantum Mechanics} (Cambridge University Press) (1988).

\bibitem{Nha} H. Nha and H. J. Carmichael, \prl {\bf 93}, 020401 (2004);  R. Garcia-Patron {\it et al.}, {\it ibid.} {\bf 93}, 130409 (2004).

\bibitem{Banaszek1999} K. Banaszek and K. W{\'o}dkiewicz, Phys. Rev. Lett. \textbf{82}, 2009 (1999).

\bibitem{Kim} J. Li, T. Fogarty, C. Cormick, J. Goold, T. Busch, and M. Paternostro, Phys. Rev. A 84, 022321 (2011);  S. -W. Lee, M. Paternostro, J. Lee, and H. Jeong, Phys. Rev. A 87, 022123 (2013); H.-J. Kim, J. Kim, and H. Nha, \pra {\bf 88}, 032109 (2013).

\bibitem{Adesso2014} G. Adesso and S. Piano, Phys. Rev. Lett. \textbf{112}, 010401 (2014).



\bibitem{Lloyd} S. Lloyd and S. L. Braunstein, Phys. Rev. Lett. {\bf 82}, 1784 (1999).

\bibitem{Eisert} J. Eisert, S. Scheel, and M. B. Plenio, Phys. Rev. Lett. {\bf 89}, 137903 (2002); J. Fiurášek, Phys. Rev. Lett. {\bf 89}, 137904 (2002); G. Giedke and J. Ignacio Cirac, Phys. Rev. A {\bf 66}, 032316 (2002).

\bibitem{Niset} J. Niset, J. Fiurášek, and N. J. Cerf, Phys. Rev. Lett. {\bf 102}, 120501 (2009).


\bibitem{Filip2014} R. Filip and L. Mišta, Jr., Phys. Rev. Lett. {\bf 106}, 200401 (2011);
M. Ježek, I. Straka, M. Mičuda, M. Dušek, J. Fiurášek, and R. Filip, Phys. Rev. Lett. {\bf 107}, 213602 (2011); 
I. Straka {\it et al.} \prl {\bf 113}, 223603 (2014). 

\bibitem{negativity} In some of quantum tasks like quantum computation and nonlocality test, the negativity in phase space, not simply non-Gaussianity, is identified as essential. See also A. Mari and J. Eisert, \prl {\bf 109}, 230503 (2012). 

\bibitem{Genoni2007} M. G. Genoni, M. G. A. Paris, and K. Banaszek, Phys. Rev. A \textbf{76}, 042327 (2007); M. G. Genoni, M. G. A. Paris, and K. Banaszek, {\it ibid.} \textbf{78}, 060303(R) (2008).



\bibitem{CHSH} J. F. Clauser, M. A. Horne, A. Shimony, and R. A. Holt, Phys. Rev. Lett. {\bf 23}, 880 (1969).

\bibitem{note} More explicitly, for a coherent state $|\alpha\rangle$, we have
\begin{eqnarray}
\mathcal{J}=&&e^{- 2 ( x_0 - \alpha_{x} )^{2}} (e^{- 2 ( y_0 - \alpha_{y} )^{2}}+e^{- 2 ( y_1 - \alpha_{y} )^{2}})\nonumber\\&& +e^{- 2 ( x_1 - \alpha_{x} )^{2}} (e^{- 2 ( y_0 - \alpha_{y} )^{2}}-e^{- 2 ( y_1 - \alpha_{y} )^{2}}).\nonumber
\end{eqnarray}
 (i) If $e^{- 2 ( y_0 - \alpha_{y} )^{2}}\ge e^{- 2 ( y_1 - \alpha_{y} )^{2}}$, using $0\le e^{-A^2}\le1$, we have $0\le\mathcal{J}\le (e^{- 2 ( y_0 - \alpha_{y} )^{2}}+e^{- 2 ( y_1 - \alpha_{y} )^{2}}) +(e^{- 2 ( y_0 - \alpha_{y} )^{2}}-e^{- 2 ( y_1 - \alpha_{y} )^{2}})=2e^{- 2 ( y_0 - \alpha_{y} )^{2}}\le2$, where the upper bound can be achieved at, e.g., $x_0=x_1= \alpha_{x}$ and $y_0=\alpha_{y}\ne y_1$. 
(ii) If $e^{- 2 ( y_0 - \alpha_{y} )^{2}}\le e^{- 2 ( y_1 - \alpha_{y} )^{2}}$, we have
$-1\le-(e^{- 2 ( y_1 - \alpha_{y} )^{2}}-e^{- 2 ( y_0 - \alpha_{y} )^{2}})\le\mathcal{J}\le e^{- 2 ( y_0 - \alpha_{y} )^{2}}+e^{- 2 ( y_1 - \alpha_{y} )^{2}}\le2$, where the lower bound -1 can be achived at, e.g., $x_0=y_0\rightarrow\infty$, $x_1= \alpha_{x}$ and $y_1= \alpha_{y}$. Therefore, from (i) and (ii), we have $-1\le\mathcal{J}\le2$ for any coherent state. 

\bibitem{Supple} See Supplemental Material for (A) a proof of the maximum Gaussian bound and an optimal set of points for test of a Gaussian state (B) a nonclassicality test using three phase-space points (C) discussion on the violation of the lower bound -1 (D) maximal J=4 of non-Gaussian states and (E) experimental details, which includes Ref. [23,27,28,33].

\bibitem{Cirelson1980} B. Cirel’son, Lett. Math. Phys. 4, \textbf{93} (1980).

\bibitem{Jeong2003} H. Jeong, W. Son, M. S. Kim, D. Ahn, and C. Brukner, Phys. Rev. A \textbf{67}, 012106 (2003).

\bibitem{Genoni2013} M. G. Genoni, M. L. Palma, T. Tufarelli, S. Olivares, M. S. Kim, and M. G. A. Paris, Phys. Rev. A \textbf{87}, 062104 (2013).

\bibitem{Hughes2014} C. Hughes, M. G. Genoni, T. Tufarelli, M. G. A. Paris, and M. S. Kim, Phys. Rev. A \textbf{90}, 013810 (2014).


\bibitem{KKim2014} S. An {\it et al.}, Nat. Phys. {\bf 11}, 193 (2015).

\bibitem{WWineland96} D. Leibfried, D. M. Meekhof, B. E. King, C. Monroe, W. M. Itano, and D. J. Wineland, Phys. Rev. Lett. \textbf{77}, 4281 (1996).

\bibitem{nha1} In the parallelogram test, a value $\mathcal{J}$ in the range of $2<\mathcal{J}<8/3^{9/8}\approx2.32$ does not represent nonclassicality of the state, unlike the case of rectangle test. This is because that the former is essentially a method of applying squeezing operation to a given state that can transform a classical state to a nonclassical one. Thus, only a value $\mathcal{J}>8/3^{9/8}\approx2.32$ is meaningful as manifesting genuine non-Gaussianity, which is also a clear signature of nonclassicality of the state by ruling out the mixture of Gaussian states including coherent states.

\bibitem{nha2} Our experimental data is collected following the same experimental sequences for both component states $\ket{0}$ and $\ket{2}$, particularly using the same set of four phase-space displacements, which makes a valid sense of ensemble average of the state $f \ketbra{0}{0} + (1-f) \ketbra{2}{2}$.



\bibitem{Laiho2010} K. Laiho, K. N. Cassemiro, D. Gross, and C. Silberhorn, Phys. Rev. Lett. \textbf{105}, 253603 (2010).

\bibitem{Silberhorn} G. Harder, D. Mogilevtsev, N. Korolkova, and Ch. Silberhorn, \prl {\bf 113}, 070403 (2014).

\bibitem{Jezek01} J. \u{R}eh\'{a}\u{e}k and Z. Hradil and M. Jez\u{e}k, Phys. Rev. A \textbf{63}, 040303(R) (2001).

\end{thebibliography}
\end{document}